\documentclass[12pt]{iopart}
\usepackage{graphicx}

\begin{document}

\title[Why a quantum state does not represent an element of physical reality]{QBism
  and the Greeks: why a quantum state does not represent an element of physical
  reality}

\author{Christopher A. Fuchs$^{1,2}$ and R\"udiger Schack$^{1,3}$}

\address{$^1$ Stellenbosch Institute for Advanced Study (STIAS),
Wallenberg Research Centre at Stellenbosch University, Marais Street, Stellenbosch 7600,
South Africa
}

\address{$^2$ Raytheon BBN Technologies, 10 Moulton Street, Cambridge MA 02138,
  USA}

\address{$^3$ Department of Mathematics, Royal Holloway, University of London,
Egham, Surrey TW20 0EX, United Kingdom}

\begin{abstract}
  In QBism (or Quantum Bayesianism) a quantum state does not represent
  an element of physical reality but an agent's personal probability assignments, reflecting
  his subjective degrees of belief about the future content of his
  experience.  In this paper, we contrast QBism with hidden-variable accounts
  of quantum mechanics and show the sense in which QBism explains quantum
  correlations. QBism's agent-centered worldview can be seen as a development
  of ideas expressed in Schr\"odinger's essay ``Nature and the Greeks''.
\end{abstract}


\section{Introduction}

In 1964 John Bell derived the inequalities which now bear his name and showed
that they are violated by quantum mechanics \cite{Bell64}. He thus established
that quantum mechanics does not admit a local hidden variable account. Here and
throughout this paper, the term ``hidden variable'' includes any mathematical
object that represents an element of physical reality and determines the
outcomes of experiments or their probabilities.

The assumption of locality thus rules out any hidden variable interpretation of
quantum mechanics. Locality also rules out directly any interpretation that
regards the quantum state as representing an element of physical reality. This
can be seen by adapting an argument by Einstein (see \cite{Caves02} and the detailed recent
discussion in \cite{Harrigan}).  Consider a maximally
entangled pair of particles far removed from each other. According to quantum
theory, by making measurements on one of the particles, an experimenter can
choose whether the state for the other particle belongs to one or the other of
two nonoverlapping sets of states (this is sometimes called ``steering''). An
interpretation that regards the quantum state as representing an element of
physical reality would thus be nonlocal, because this element of physical
reality could be manipulated at a distance.

Recently, Pusey, Barrett, and Rudolph (PBR) \cite{PBR}, Colbeck and Renner (CR)
\cite{Colbeck12}, and other authors, showed that, under certain conditions, in
any hidden variable theory the quantum state must be a function of the hidden
variables. These papers are set in the ``ontological model'' framework
introduced by Harrigan and Spekkens \cite{Harrigan}, where 
a distinction is made between ``psi-ontic'' and ``psi-epistemic''
theories---ones where the quantum states are included among the
states of reality and ones where quantum states are simply subjective states of
knowledge. (For a recent discussion on the relation between PBR and arguments
based on nonlocality see, e.g., \cite{Emerson}.)
In the wider community and particularly the blogosphere, the
results of PBR and CR are often naively taken to imply the death of the very
notion of epistemic quantum states.  But, it must not be forgotten that the
psi-ontic versus psi-epistemic distinction as defined in those papers is
between two kinds of hidden variable theories, and that the existence of hidden
variables is not at all implied by quantum mechanics.

QBism \cite{Caves02,Fuchs10,Fuchs13,FMS} is an explicitly local interpretation
of quantum mechanics in which there is no room for hidden variables. According
to QBism, quantum mechanics is a tool any agent can use to evaluate his
probabilistic expectations for his personal experience. A quantum state does
not represent an element of physical reality external to the agent, but
reflects the agent's personal degrees of belief about the future content of his
experience.

In Section \ref{sec:Greeks}, we show how the agent-centered worldview of QBism
arises naturally as a development of the fundamental issues identified by
Schr\"odinger in his essay ``Nature and the Greeks'' \cite{Schroedinger1951}.
Section \ref{sec:qbism} gives a short overview of QBism. Section
\ref{sec:BornRule} discusses the function of the Born rule in QBism and contrasts
QBism with hidden variable theories. In Section \ref{sec:explanation} we show
the sense in which QBism explains quantum correlations, and Section \ref{sec:conclusion}
concludes the paper.

\section{QBism and the Greeks}  \label{sec:Greeks}

In the essay ``Nature and the Greeks'' \cite{Schroedinger1951},
Schr\"odinger writes: ``Gomperz says
[...] that our whole modern way of thinking is based on Greek thinking; it is
therefore something special, something that has grown historically over many
centuries, {\it not\/} the general, the only possible way of thinking about
Nature. He sets much store on our becoming aware of this, of recognising the
peculiarities as such, possibly freeing us from their well-nigh irresistible
spell.''

Schr\"odinger singles out two fundamental features of modern science
that are influenced by Greek thinking in this way. One is ``the assumption that
the world can be understood.'' The other is ``the simplifying provisional
device of excluding the person of the `understander' (the subject of
cognizance) from the rational world-picture that is to be constructed.''

About the first of these, Schr\"odinger remarks that ``one would in this
context have to discuss the questions: what does comprehensibility really mean,
and it what sense, if any, does science give explanations?'' The question of
explanation is often brought up in discussions of the foundations of quantum
mechanics. Derivations of the Bell inequalities have been phrased in terms of
possible explanations of the correlated data produced in a Bell experiment
\cite{bell81}. We will address the issue of explanation from a QBist
perspective in Section \ref{sec:explanation}.

Here we focus on the second special feature of modern science
identified by Schr\"odinger, namely that ``the scientist subconsciously, almost
inadvertently, simplifies his problem of understanding Nature by disregarding
or cutting out of the picture to be constructed himself, his own personality,
the subject of cognizance.'' According to Schr\"odinger, this ``leaves gaps,
enormous lacunae, leads to paradoxes and antinomies whenever, unaware of this
initial renunciation, one tries to find oneself in the picture, or to put
oneself, one's own thinking and sensing mind, back into the picture.''

An example of this fundamental difficulty is provided by the quantum
measurement problem: How is it that an agent experiences a single outcome when
he performs a measurement on a system in a superposition state? In most
accounts of the measurement problem, the quantum state is regarded as
agent-independent and objective, and hence as belonging to what Schr\"odinger
calls the rational world-picture from which the subject of cognizance is
excluded. The measurement problem can be seen as a symptom of the ``paradoxes
and antinomies'' that one finds when one tries to connect this world-picture to
the experience of an agent. The many decades of ultimately unsuccessful
attempts to resolve the measurement problem attest to its fundamental nature.

The following fragment, quoted twice in Schr\"odinger's essay, shows
that the core of the problem was clearly understood by Democritus:
``(Intellect:) Sweet is by convention, and bitter by convention, hot by
 convention, cold by convention, colour by convention; in truth there are but
 atoms and the void.
(The Senses:) Wretched mind, from us you are taking the evidence by which you
 would overthrow us? Your victory is your own fall.''
Schr\"odinger comments ``You simply cannot put it more briefly and clearly.''

There is no measurement problem in QBism because the agent and the agent's
experience are part of the story from the beginning. QBism is thus breaking
free from the ``irresistible spell'' of Greek thinking and abandons ``the
simplifying provisional device of excluding the person of the `understander'
(the subject of cognizance) from the rational world-picture that is to be
constructed.'' The next section will give details of this move. Both the
locality assumption discussed in the introduction and a contemporary
understanding of probability provide strong motivations for this move,
independently of Schr\"odinger's views on the impact of Greek thinking on the
presuppositions of contemporary science.

\section{QBism}
\label{sec:qbism}

The fundamental primitive of QBism is the concept of experience. According
to QBism, quantum mechanics is a theory that any agent can use to evaluate his
expectations for the content of his personal experience.

QBism adopts the personalist Bayesian probability theory pioneered by Ramsey
\cite{Ramsey26} and de Finetti \cite{DeFinetti31} and put in modern form by
Savage \cite{Savage54} and Bernardo and Smith \cite{Bernardo94} among
others. This means that QBism interprets all probabilities, in particular those
that occur in quantum mechanics, as an agent's personal, subjective degrees of
belief. This includes the case of certainty---even probabilities 0 or 1 are
degrees of belief \cite{Caves07}. Probabilities acquire an operational meaning
through their use in decision making, or gambling: an agent's probabilities are
defined by his willingness to place or accept bets on the basis of those
probabilities. In this framework, the usual probability rules can be derived
from the requirement that an agent's probability assignments should not lead to a
sure loss in a single instance of a bet, a requirement known as Dutch-book
coherence. The probability rules are therefore of a normative character.

Dutch-book coherence for one agent does not put any constraints on another
agent's probability assignments. The set of probabilities used by an agent have
validity for that agent only. The general theory---degrees of belief
constrained by Dutch book coherence---can be used by any agent.  But one cannot
mix the probability assignments made by different users of the theory.

In QBism, a measurement is an action an agent takes to elicit an
experience. The measurement outcome is the experience so elicited. The
measurement outcome is thus personal to the agent who takes the measurement
action. In this sense, quantum mechanics, like probability theory, is a single
user theory. A measurement does not reveal a pre-existing value. Rather, the
measurement outcome is created in the measurement action.

According to QBism, quantum mechanics can be applied to any physical
system. QBism treats all physical systems in the same way, including atoms,
beam splitters, Stern-Gerlach magnets, preparation devices, measurement
apparatuses, all the way to living beings and other agents. In this, QBism
differs crucially from various versions of the Copenhagen interpretation. A common thread among those instead is that measuring and preparation devices, in their operation as such, must be treated as belonging to a separate classical domain outside the scope of quantum mechanics \cite[Sect.\ 3]{Peres2002}.

An agent's beliefs and experiences are necessarily local to that agent. This
implies that the question of nonlocality simply does not arise in QBism. QBist
quantum mechanics is local because, for any user of quantum mechanics,
quantum states encode the user's personal degrees of belief for the contents of
his own experience \cite{FMS}.

Quantum states are represented by density operators $\rho$ in a Hilbert space
assumed to be finite dimensional. A measurement (an action taken by the agent)
is described by a POVM $\{F_j\}$, where $j$ labels the potential outcomes
experienced by the agent. The agent's personalist probability $q(j)$ of
experiencing outcome $j$ is given by the Born rule,
\begin{equation}q(j)={\rm tr}(\rho F_j) \;. \label{bornRule}\end{equation}
Similar to the probabilities on the left-hand side of the Born rule, QBism
regards the operators $\rho$ and $F_j$ on the right-hand side as judgements made
by the agent, representing his personalist degrees of belief.

\section{The function of the Born rule}  \label{sec:BornRule}

The Born rule as written in Eq.~(\ref{bornRule}) appears to connect
probabilities on the left-hand side of the equation with other kinds of
mathematical objects---operators---on the right-hand side. It turns out to be
possible, however, to rewrite the rule entirely in terms of probabilities
\cite{Fuchs10,Fuchs13}.
For this, consider the scenario of Figure 1, where a reference measurement is
introduced in order to characterize both the system state $\rho$ and the
POVM $\{F_j\}$.

\begin{figure}
\begin{center}
\includegraphics[scale=0.5]{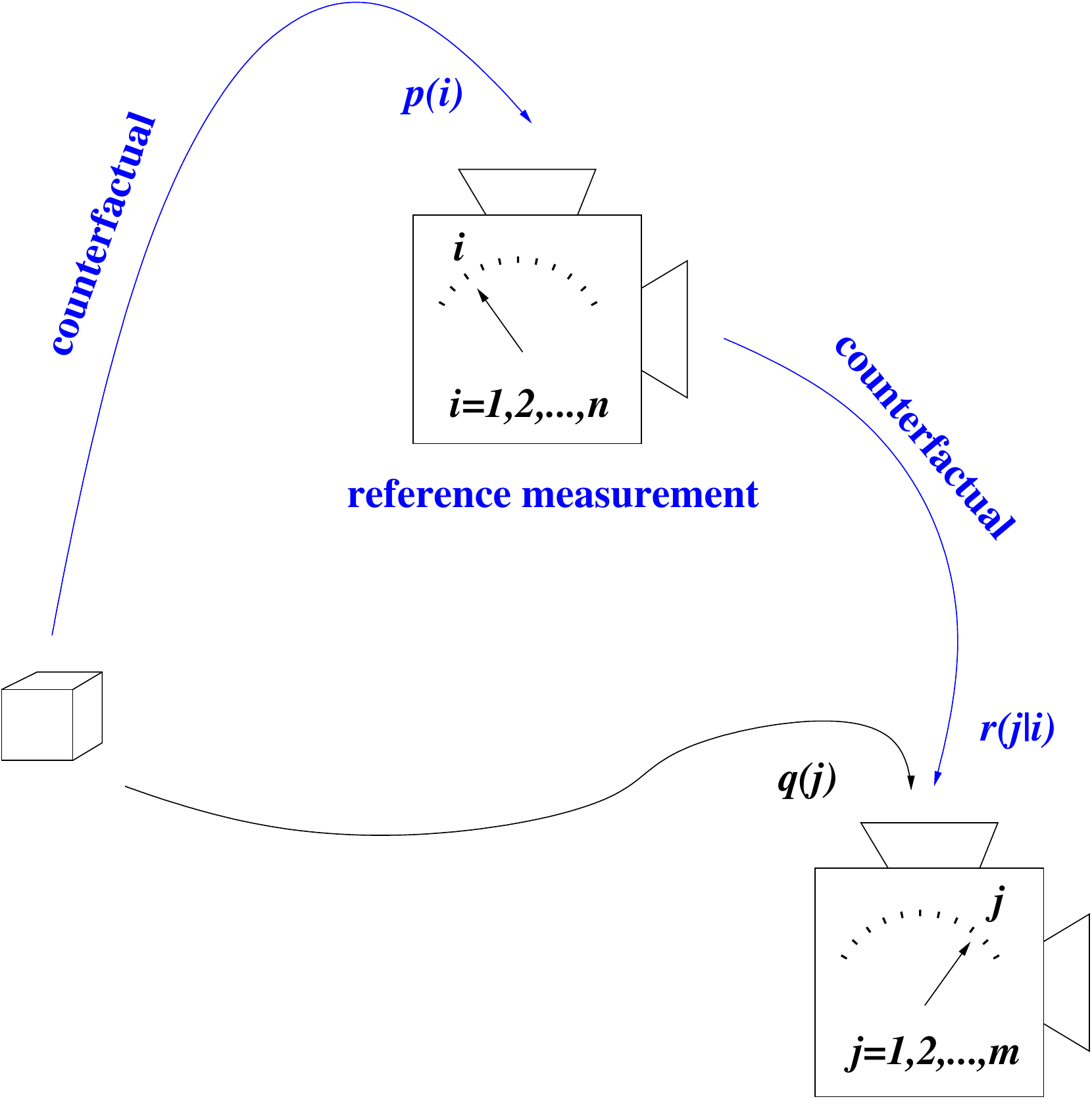}
\end{center}
\caption{Analysing a measurement in an agent-centered way: the index $j$
  labels the outcomes of some actual measurement the agent intends to perform,
  and $i$ labels the outcomes of a reference measurement which the
  agent might perform but which remains counterfactual. In both classical
  mechanics and quantum mechanics there exist such reference measurements for
  which the agent's probabilities $q(j)$ for outcome $j$ can be expressed in
  terms of his probabilities $p(i)$ for outcome $i$ and his conditional
  probabilities $r(j|i)$ for outcome $j$ given outcome $i$.}
\end{figure}

We assume that the agent's reference measurement is an arbitrary
informationally complete POVM, $\{E_i\}$, such that each $E_i$ is of rank 1,
i.e., is proportional to a one-dimensional projector $\Pi_i$. Such measurements
exist for any finite Hilbert-space dimension. Furthermore, we assume that, if
the agent carries out the measurement $\{E_i\}$ for an initial state $\rho$,
upon getting outcome $E_i$ he would update to the post-measurement state
$\rho_i=\Pi_i\rho\Pi_i/{\rm tr}(\rho\Pi_i)$. Because the reference measurement
is informationally complete, any state $\rho$ corresponds to a
unique vector of probabilities $p(i)={\rm tr}(\rho E_i)$, and any POVM
$\{F_j\}$ corresponds to a unique matrix of conditional probabilities
$r(j|i)={\rm tr}(F_j\Pi_i)$.

The operators $\rho$ and $F_j$ on the right-hand side of the Born rule are thus
mathematically equivalent to sets of probabilities $p(i)$ and conditional
probabilities $r(j|i)$, respectively. In this sense, POVMs as well as quantum
states {\it are\/} probabilities. In QBism, POVMs as well as quantum states
represent an agent's personal degrees of belief. The Born rule then
becomes
\begin{equation}
q(j)=f\Big(\{p(i)\},\{r(j|i)\}\Big)  \;,   \label{eq:f}
\end{equation}
where the precise form of the function $f$ depends on the details of the
reference measurement. The Born rule allows the agent to calculate his outcome
probabilities $q(j)$ in terms of his probabilities $p(i)$ and $r(j|i)$ defined
with respect to a counterfactual reference measurement.

In QBism, the Born rule functions as a coherence requirement. Rather than
setting the probabilities $q(j)$, the Born rules relates them to those defining
the state $\rho$ and the POVM $\{F_j\}$.  Just like the standard rules of
probability theory, the Born rule is normative: the agent ought to assign
probabilities that satisfy the constraints imposed by the Born rule. Unlike the standard rules
of probability theory however, which can be derived from Dutch-book coherence alone,
the Born rule is empirical. It is a statement about the physical world.

We will now show that the scenario of Figure 1 captures not only the essential
difference between classical physics and quantum theory, but also the essential
difference between QBism and hidden variables theories.  Classical physics
rests on the assumption that, for every system, there exists a reference
measurement such that, for every actual measurement, the following
holds. As before, let
$p(i)$ denote an agent's probabilities for outcome $i$ in the reference
measurement (the $p(i)$ characterize the agent's system state), let $r(j|i)$
denote his probabilities for outcome $j$ in the actual measurement given that
the reference measurement was carried out and resulted in outcome $i$ (in a
deterministic theory the $r(j|i)$ would be restricted to values 0 or 1), and
let $q(j)$ denote the probabilities of outcome $j$ in the actual measurement
assuming that the reference measurement remains counterfactual. Then
\begin{equation}  \label{eq:totalProb}
q(j)=\sum_i p(i) r(j|i) \;.
\end{equation}
Since in the definition of $q(j)$, the reference measurement remains
counterfactual, Eq.~(\ref{eq:totalProb}) is not implied by probability theory
\cite{Fuchs13}. It is a physical postulate. This formulation of the classical
postulate is agent-centered. It connects an agent's degrees of belief about the
outcomes of the reference measurement with his degrees of belief about the
outcomes of the actual measurement.

The agent (or subject) might be thought to be removable from the picture by taking the variables
$i$ to represent external states of reality that determine the probabilities $r(j|i)$. In this
case $p(i)$ denotes the probability that the state of reality is $i$. The
central assumption of classical physics now takes the form that, in principle,
there is a measurement that simply reads off the value of $i$. The
classical law Eq.~(\ref{eq:totalProb}) is then a consequence of
probability theory. It is the same equation as before, but it now refers to an
agent-independent reality. In a nutshell, this is the 2000 year old Greek
maneuver identified by Schr\"odinger that excluded the subject from the world
picture.

Of course the world is not classical. There is in general no reference measurement
such that the classical law Eq.~(\ref{eq:totalProb}) holds. QBism takes this
fact---the nonexistence of such a reference measurement---as an expression of
the idea that the subject cannot be removed from the world picture.

By contrast, ontological models \cite{Harrigan}, or hidden variable models, try
to preserve the concept of an agent-independent reality. Similar to the
agent-independent formulation of classical physics, they analyze measurements
in terms of an external state of reality $i$, a probability distribution $p(i)$ over
external states of reality, and a conditional probability distribution $r(j|i)$ which
gives the probability for outcome $j$ for each external state of reality $i$.
Furthermore, the classical law Eq.~(\ref{eq:totalProb}) is assumed to
hold. Since quantum mechanics rules out an interpretation of $i$ as the outcome
of a reference measurement, in ontological models Eq.~(\ref{eq:totalProb}) does
not follow from probability theory but is an independent postulate. In these
models, the Born rule is either a further independent postulate or follows from
further assumptions about the variables $i$.

QBism keeps the idea of a reference measurement and thus keeps the subject in
the center. Since the reference measurement is assumed to remain
counterfactual, probability theory alone has nothing to say about the relation
between the probabilities $q(j)$, $p(i)$ and $r(j|i)$. The Born rule can thus be
seen as an addition to probability theory, a normative requirement of {\it
  quantum Bayesian coherence\/} \cite{Fuchs13}, which applies whenever the
agent contemplates a particular kind of reference measurement. The functional
relationship Eq.~(\ref{eq:f}) depends on the details of the reference
measurement. In the special case that the reference measurement is a symmetric
informationally complete POVM (SIC) \cite{Zauner99,Caves99,Renes04},
Eq.~(\ref{eq:f}) takes the simple form \cite{Fuchs10,Fuchs13}
\begin{equation}  \label{eq:sic}
q(j)=\sum_i \left((d+1)p(i)-\frac1d\right) r(j|i) \;.
\end{equation}
The authors have conjectured that this form of the Born rule may be used as an
axiom in a derivation of quantum theory
\cite{Fuchs13,FuchsSchack2011,Appleby2014}. 

Indeed recently there have been several information-theoretic axiomatic
derivations of quantum theory \cite{Chiribella2011,Masanes2011,Torre2012}. These
may provide important clues and techniques for how to proceed to a full
derivation of quantum theory from Eq.~(\ref{eq:sic}), which so far has not been
complete.  The key question that remains is in identifying what minimal further
principles must be added to Eq.~(\ref{eq:sic}) for the project to be
successful.  What would be unique about this approach, if it proves successful,
is the way it would pull the scenario depicted in Figure 1 to the front and
center of the mathematical structure of quantum theory.  In
Ref.~\cite{FuchsStacey2014}, one of us (CAF) expands on why this notion is
considered key for a thorough-going QBist expression of quantum theory. In a
nutshell, it is that Eq.~(\ref{eq:sic}) gives quantitative expression to the
idea that the agent cannot be removed from the world picture.

\section{Explanation}
\label{sec:explanation}

According to QBism, the quantum formalism is an addition to probability
theory (see the previous section). One should therefore expect that
explanations offered by quantum theory have a similar character to explanations
offered by probability theory.

Here is a simple example from probability theory. Assume an agent's prior
probabilities for a coin tossing experiment are such that for him the coin
tosses are independent and Heads and Tails are equally likely in each toss. The
agent now considers tossing the coin 100 times, denoting by $h$ the number of
Heads. Using simple properties of the binomial distribution, the agent expects
$h$ to lie between 30 and 70 with probability close to 1. When he performs
the experiment, he happens to find the value $h=57$.

Probability theory explains the agent's expectations. The theory allows the
agent to understand why, given his prior, he should be almost certain that he
will find a value of $h$ between 30 and 70. On the other hand, probability
theory does not provide any explanation for why the agent found the particular
value $h=57$. This hardly limits the wide-ranging explanatory power of
probability theory as witnessed by any standard probability text.

Our second example is quantum mechanical. Consider an experimenter who prepares
a spin-1/2 particle in an $x$ eigenstate and performs a measurement using a
Stern-Gerlach device oriented along the $z$ axis. This setup encodes the
experimenter's prior. Given this prior, quantum mechanics explains why, in
order to be coherent, the experimenter should assign probability 1/2 to each of
the two possible outcomes. Say the experimenter experiences the outcome
``up''. Quantum mechanics does not explain why he experiences ``up'' and not
``down''. Far from being a limitation of the theory, this is an expression of
the QBist idea that neither the outcome of the measurement nor its probability
are determined by some hidden variables: measurement outcomes or their
probabilities are not a function solely of the physical reality external to the agent.
The explanations provided by quantum mechanics are exactly those one would
hope for in a world in which measurements are acts of creation, i.e., in a
world that is unfinished and open.

The above example extends naturally to the case of repeated measurements.
Assuming the experimenter has an appropriate prior, quantum mechanics explains
why he should expect, with probability close to 1, that in many repetitions of
the spin measurement the proportion of spin-up outcomes he experiences will be
close to 1/2. Quantum mechanics does not provide any explanation for the
particular proportion the agent finds---this is just as it was with the coin toss example. Yet, even more than in the case of
probability theory, this does not prevent quantum mechanics from having
unprecedented explanatory power.

Correlations are just a special case of more general probability
assignments. To explain a correlation is therefore no different than to explain
a probability assignment.  Here is an example for how correlations arise in
quantum mechanics. Suppose that an agent considers performing a measurement on
a spin-3/2 particle. For given labels $(a,b)$, the measurement is
assumed to be of the form $\{A^a_x\otimes B^b_y\}$, where the $A^a_x$ and
$B^b_y$ correspond to projection operators onto two complementary
(two-dimensional) sub-algebras of the full four-dimensional Hilbert space. If
we denote the agent's prior state for the particle by $|\psi\rangle$, the
agent's probabilities $p(x,y|a,b)$ for experiencing the outcome $(x,y)$ if he
chooses to enact the measurement labeled by $(a,b)$ are given by
\begin{equation}   \label{pxyab}
p(x,y|a,b)=\langle\psi| A^a_x\otimes B^b_y|\psi\rangle \;.
\end{equation}
In this way the quantum formalism explains why, given the agent's prior beliefs,
he ought to assign the correlations $p(x,y|a,b)$.

If the agent performs measurements of this type on a large number $n$ of
particles for which his prior is the product state $|\psi\rangle^{\otimes n}$,
he can record the frequencies with which the different outcomes occur for each
setting $(a,b)$ in a data table, $d(x,y|a,b)$. As before, quantum mechanics
explains why the agent should expect the measured frequencies to lie in a
certain range, but does not provide an explanation for the particular numbers
the agent obtains in a given realization of the data table.

The above considerations remain unchanged in the case that the correlations
$p(x,y|a,b)$ implied by the prior state and measurement operators violate a
Bell inequality. Of course, Bell inequalities are not usually introduced for
sub-algebras of a spin-3/2 particle, but for measurements on two space-like
separated subsystems.  In QBism, however, there is no important conceptual
difference between these two situations.

The above considerations also remain unchanged in the case of perfect
correlations, $p(x,y|a,b)\in\{0,1\}$. Even these are an agent's personal
probabilities for his future experiences. QBism treats all quantum systems and
all measurements on an equal footing. That unperformed measurements have no
outcomes is true for all measurements, independently of whether or not the
agent assigns probability 1 to one of the outcomes. A statement such as
$p(y=0)=1$ expresses the agent's personal belief that the measurement outcome
will be $y=0$, a belief that is given a quantitative expression through the
bets he would accept on this outcome---here he would bet an arbitrary amount
against the promise of an arbitrarily small gain.  It has been argued by
Timpson \cite{Timpson08} that it might be irrational for an agent to make a
probability assignment such as $p(y=0)=1$ unless the agent also believed in the
existence of a ``truth maker'' that guarantees that the outcome will indeed be
$y=0$. Timpson's argument would lead to the introduction of an additional
constraint on the assignment on probabilities, beyond the constraints imposed
by the probability calculus and, via the Born rule, quantum mechanics. Such an
extra constraint is not implied by quantum theory and ultimately amounts to the
introduction of hidden variables. It is therefore ruled out by the QBist view
of the world \cite[pp.~1809--1810 and links therein]{Struggles}.

\section{Summary}   \label{sec:conclusion}

According to QBism, quantum mechanics is a theory any agent can use to
more safely gamble on his potential future experiences. Quantum mechanics permits any agent to
quantify, on the basis of his past experiences, his probabilistic expectations
for his future experiences. QBism takes measurement outcomes as well as quantum
states to be personal to the agent using the theory. In QBism, there are no
agent-independent elements of physical reality that determine either
measurement outcomes or probabilities of measurement outcomes. Rather, every
quantum measurement is an action on the world by an agent that results in the
creation of something entirely new. QBism holds this to be true not only for
laboratory measurements on microscopic systems, but for any action an agent
takes on the world to elicit a new experience. It is in this sense that agents
have a fundamental creative role in the world.


\section*{References}
\begin{thebibliography}{10}

\bibitem{Bell64}
J. S. Bell, ``On the Einstein-Podolsky-Rosen Paradox,'' Physics {\bf 1}, 195 (1964).

\bibitem{Caves02}
C. M. Caves, C. A. Fuchs, and R. Schack, ``Quantum Probabilities as Bayesian Probabilities,''
Phys.\ Rev. A {\bf 65}, 022305 (2002).

\bibitem{Harrigan}
N. Harrigan and R. W. Spekkens,
``Einstein, Incompleteness, and the Epistemic View of Quantum States,''
Found.\ Phys.\ {\bf 40}, 125 (2010).

\bibitem{PBR}
M. F. Pusey, J. Barrett, and T. Rudolph,
``On the Reality of the Quantum State,''
Nature Phys.\ {\bf 8}, 475 (2012).

\bibitem{Colbeck12}
R. Colbeck and R. Renner,
``Is a system's wave function in one-to-one correspondence with its elements of reality?''
Phys.\ Rev.\ Lett.\ {\bf 108}, 150402 (2012).

\bibitem{Emerson}
J. Emerson, D. Serbin, C. Sutherland, and V. Veitch,
``The whole is greater than the sum of the parts:\ on the possibility of purely
statistical interpretations of quantum theory,''
{\tt arXiv:1312.1345} (2013).

\bibitem{Fuchs10}
C. A. Fuchs, ``QBism, the Perimeter of Quantum Bayesianism,''
{\tt arXiv:1003.5209} (2010).

\bibitem{Fuchs13}
C. A. Fuchs and R. Schack, ``Quantum-Bayesian Coherence,''
Rev.\ Mod.\ Phys.\ {\bf 85}, 1693 (2013).

\bibitem{FMS}
C. A. Fuchs, N. D. Mermin, and R. Schack,
``An Introduction to QBism with an Application to the Locality of Quantum Mechanics,'' Am.\ J.\ Phys.\ {\bf 82}, 749 (2014).

\bibitem{Schroedinger1951} E. Schr\"odinger, {\it `Nature and the Greeks' and
  `Science and Humanism'} (Cambridge University Press, Cambridge, UK, 1951).

\bibitem{bell81}
J. S. Bell, ``Bertlmann's socks and the nature of reality,''
Journal de Physique Colloques {\bf 42}, C2-41 (1981).

\bibitem{Ramsey26}
F. P. Ramsey ``Truth and Probability,'' in F.~P. Ramsey, {\sl The Foundations of Mathematics and other Logical Essays}, edited by R.~B. Braithwaite (Harcourt, Brace and Company, New York, 1931), p.~156.

\bibitem{DeFinetti31}
B. de~Finetti, ``Probabilismo,'' Logos {\bf 14}, 163 (1931); transl., ``Probabilism,'' Erkenntnis {\bf 31}, 169 (1989).

\bibitem{Savage54}
L.~J. Savage, {\sl The Foundations of Statistics} (John Wiley \& Sons, New York, 1954).

\bibitem{Bernardo94}
J.~M. Bernardo and A.~F.~M. Smith, {\sl Bayesian Theory} (Wiley, Chichester, 1994).

\bibitem{Caves07}
C.~M. Caves, C.~A. Fuchs, and R.~Schack, ``Subjective Probability and Quantum Certainty,'' Stud.\ Hist.\ Phil.\ Mod.\ Phys.\ {\bf 38}, 255 (2007).

\bibitem{Peres2002}
A. Peres, ``Karl Popper and the Copenhagen Interpretation,'' Stud.\ Hist.\ Phil.\ Mod.\ Phys.\ {\bf 33}, 23 (2002).

\bibitem{Zauner99}
G. Zauner, {\sl Quantum Designs --- Foundations of a Non-Commutative Theory of Designs} (in German), PhD thesis, University of Vienna (1999).

\bibitem{Caves99}
C. M. Caves, ``Symmetric Informationally Complete POVMs,'' posted at {\tt
  http://info. phys.unm.edu/$\sim$caves/reports/infopovm.pdf} (9 September 1999).

\bibitem{Renes04}
J. M. Renes, R. Blume-Kohout, A. J. Scott, and C. M. Caves, ``Symmetric Informationally Complete Quantum Measurements,'' J. Math.\ Phys.\ {\bf 45}, 2171 (2004).

\bibitem{FuchsSchack2011}
C. A. Fuchs and R. Schack, ``A Quantum-Bayesian Route to Quantum-State Space,''
Found.\ Phys.\ {\bf 41}, 345 (2011).

\bibitem{Appleby2014}
D. M. Appleby, C. A. Fuchs, and H. Zhu, ````Introducing the Qplex: A Dynamic Arena for Quantum Probability Vectors,'' in preparation.

\bibitem{Chiribella2011}
G. Chiribella, G. M. D'Ariano, and P. Perinotti, ``Informational Derivation of Quantum Theory,'' Phys.\ Rev.\ A, {\bf 84}, 012311 (2011).

\bibitem{Masanes2011}
L. Masanes and M. P. M\"uller, ``A Derivation of Quantum Theory from Physical Requirements,'' New J. Phys.\ {\bf 13}, 063001 (2011).

\bibitem{Torre2012}
G. de la Torre, L. Masanes, A. Short, and M. M\"uller, ``Deriving Quantum Theory from Its Local Structure and Reversibility,'' Phys.\ Rev.\ Lett.\ {\bf 109}, 090403 (2012).

\bibitem{FuchsStacey2014}
C. A. Fuchs and B. C. Stacey, ``Some Negative Remarks on Operational Approaches to Quantum Theory,'' {\tt arXiv:1401.7254} (2014).

\bibitem{Timpson08}
C. J. Timpson, ``Quantum Bayesianism:\ A Study,'' Stud.\ Hist.\ Phil.\ Mod.\ Phys.\ {\bf 39}, 579 (2008).

\bibitem{Struggles}
C. A. Fuchs, {\sl My Struggles with the Block Universe:\ Selected Correspondence, January 2001 -- May 2011}, {\tt arXiv:1405.2390}.

\end{thebibliography}
\end{document}